\documentclass[prl,preprint,showpacs,superscriptaddress,floatfix]{revtex4} 
\usepackage{graphicx,amsmath,amssymb,natbib}

\newcommand{\figwidth}{0.90\columnwidth}

\newcommand{\Fig}[1]{Fig.~\ref{#1}}
\newcommand{\Figlong}[1]{Figure~\ref{#1}}

\newcommand{\Fsqt}{ F_{\rm s}(q,t) }

\begin{document}

\title{Glassy dynamics in monodisperse hard ellipsoids}

\author{Patrick Pfleiderer}
\email{pfleider@uni-mainz.de}
\author{Kristina Milinkovic}
\author{Tanja Schilling}
\affiliation{Institut f\"ur Physik, Johannes Gutenberg-Universit\"at,
Staudinger Weg 7, D-55099 Mainz, Germany}

\date{\today}

\begin{abstract} 
We present evidence from computer simulations for glassy dynamics in
suspensions of monodisperse hard ellipsoids. In equilibrium, almost 
spherical ellipsoids show a first order transition from an isotropic
phase to a rotator phase. When overcompressing the isotropic phase
into the rotator regime, we observe super-Arrhenius slowing down of
diffusion and relaxation, accompanied by two-step relaxation in
positional and orientational correlators. The effects are strong
enough for asymptotic laws of mode-coupling theory to apply. Glassy
dynamics are unusual in monodisperse systems. Typically,
polydispersity in size or a mixture of particle species is
prerequisite to prevent crystallization. Here, we show that a slight
particle anisometry acts as a sufficient source of disorder. This
sheds new light on the question of which ingredients are required for
glass formation. 
\end{abstract}


\maketitle

Hard-particle models play a key role in statistical mechanics. 
They are conceptually and computationally simple, and they offer insight 
into systems in which particle 
shape is important, including atomic, molecular, colloidal, and granular
systems. Ellipsoids are a classic model of non-spherical particles.
We report here that this simple anisometry can hinder crystallization and
facilitate glassy dynamics -- a phenomenon which does usually not occur 
in monodisperse
systems. Typically, polydispersity, additional particle species, or other
sources of disorder are needed for the development of glass-like
behavior, i.e. drastic slowing down of the dynamics without a change
in structure.

In recent years, there have been two studies closely related to our topic, 
which we briefly summarize here: 
Letz and coworkers \cite{letz.schilling.latz:2000} have applied idealized
molecular mode-coupling theory (MMCT
\cite{franosch-pre-56-5659-1997,Schilling.Scheidsteger:1997}) to 
the hard-ellipsoid fluid. In addition to conventional mode-coupling theory
(MCT) \cite{goetzeLiqFreezGlass:1991},
MMCT takes orientational degrees of freedom into 
account. For nearly spherical ellipsoids, they predicted a discontinuous glass 
transition in positional and orientational degrees 
of freedom. MMCT locates the transition inside the coexistence
region between the isotropic fluid and the positionally ordered phases. 
In addition, a continuous transition was predicted upon further compression 
into the rotator regime. This transition 
affects only the odd-parity orientational correlators, 
e.g.~$180^{\circ}$ flips. 

However, MCT cannot make statements about crystal nucleation. 
Hence, the MCT prediction of a
glass transition is not sufficient to conclude that the transition will occur
in a simulation or experiment. 
A prominent example is the overcompression of monodisperse
hard spheres. Here, the nucleation barrier can be easily crossed, and 
crystallization always prevents glass formation.

De Michele et al.~\cite{demichele.schillingr.sciortino:2007} have
recently studied the dynamics of hard ellipsoids by molecular dynamics
simulations. 
The states which they simulated were mostly located in the
isotropic region. They computed isodiffusivity lines, which 
showed that the dynamics of the positional and orientational degrees 
of freedom were 
decoupled, since the positional isodiffusivity lines crossed the 
orientational ones at nearly $90^{\circ}$. This decoupling also appeared in
correlation functions.
The self-part of the intermediate scattering function displayed slight
stretching only when overcompressing nearly spherical
ellipsoids, while the second-order orientational correlator showed such
stretching only for sufficiently elongated particles, i.e.~near the
isotropic-nematic transition. Clear indicators
of glassy dynamics, however, would include a strong increase of
relaxation times with 
volume fraction, even pointing towards dynamical arrest. Typically,
correlators then develop a two-step decay, whose
second step is affected by this slowing down. Such phenomena were not 
seen in \cite{demichele.schillingr.sciortino:2007} as overcompression
was not significant. 

We have performed Monte Carlo (MC) and molecular dynamics (MD) 
simulations for hard symmetrical ellipsoids of length-to-width ratios 
$l/w=1.25$ (prolate) and $l/w=0.8$ (oblate). We overcompressed these systems
into the rotator regime (i.e.~beyond volume fraction $\phi \approx
0.55$ \cite{frenkel.mulder:1985}). We found
two-step relaxation both in positional
and even-parity orientational correlators. Positional and
orientational relaxation slow down more strongly than an
Arrhenius law. Odd-parity orientational correlators indicate that
flipping is not affected. The observed glassy dynamics are strong enough to
compare with MMCT. Also, we compare the MC and MD results.

The systems were equilibrated using MC at constant
particle number $N$, pressure $P$ and temperature $T$
\cite{wood:1968,mcdonald:1972} in a cubic box with periodic boundaries. 
Each system contained more than $3000$ particles. Random isotropic 
configurations
were used as starting configurations. Towards the end of each run, a
configuration with a volume fraction close to the average volume fraction was
chosen and scaled to the 
average volume fraction exactly. The systems were considered equilibrated
\footnote{For the case of overcompression, ``equilibrated'' here means 
within the metastable isotropic basin.}
when the volume fraction had settled and all positional and the orientational
correlators were independent of absolute simulation time. For production, we
used MC and MD. In the MC simulations (now at constant 
volume) the step sizes were fixed to small values, so that unphysical 
grazing moves were negligible. The particles then mimic
Brownian motion, similar to colloidal ellipsoids suspended in a liquid. The
step sizes were the same for all runs, hence a unique time scale could be
established. The MD simulations implemented free flight and elastic
collisions \cite{allen.frenkel.talbot:1989}. 

The systems had a slight tendency to crystallize to the rotator phase
at high volume fractions. To monitor crystallinity, we computed the 
local positional 
order using the bond-orientational order parameter $q_{\rm 6}$
\cite{Wolde.RuizM.Frenkel:1995}. The fraction of particles which were part of
crystalline clusters never exceeded 2.6\% and was typically below 0.5\%.


\begin{figure}
\begin{center}
\includegraphics[ 
  width=\figwidth
  ]{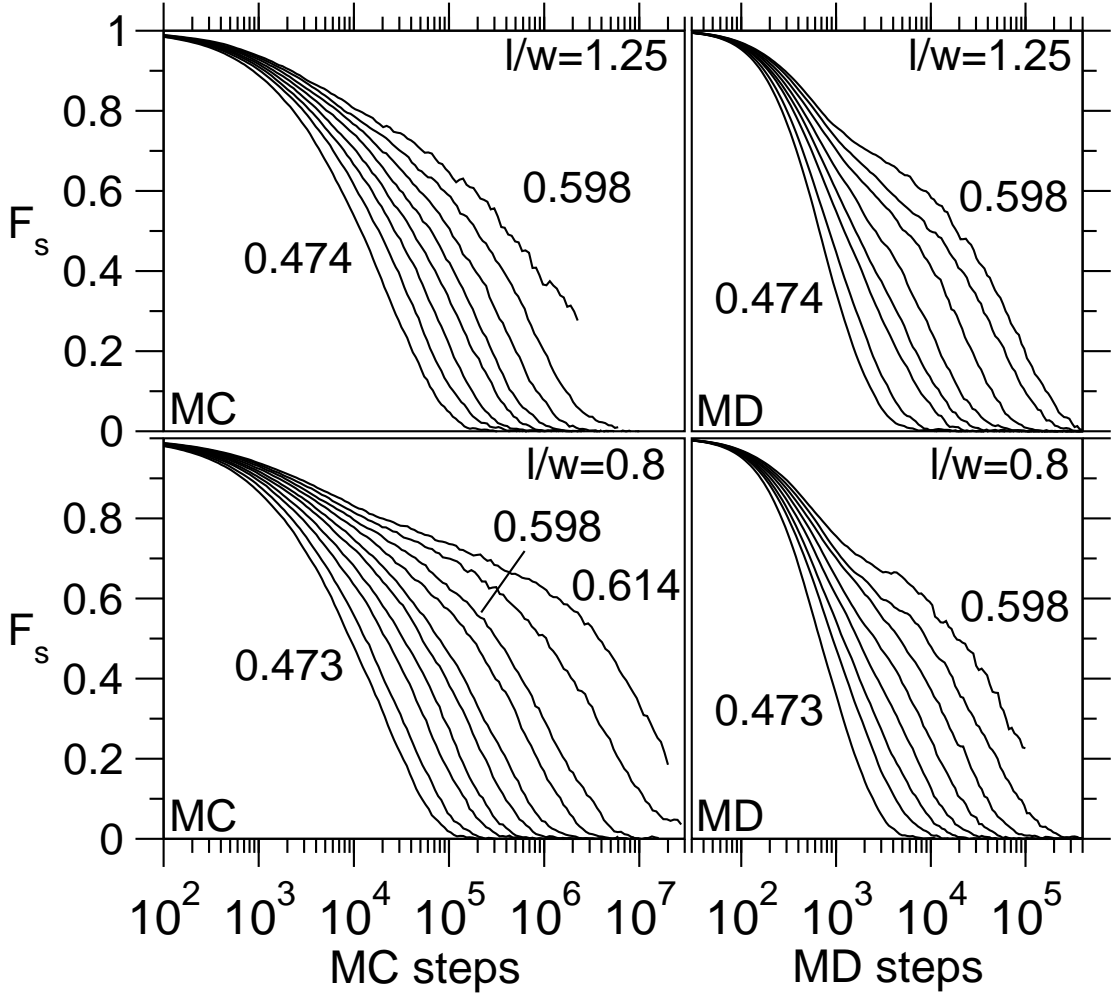}
\caption{\label{fs} Self-intermediate scattering functions at several
  volume fractions $\phi$ (prolate: 0.474, 0.511, 0.533, 0.551, 0.565,
  0.578, 0.588, 0.598; oblate: 0.473, 0.504, 0.533, 0.550, 0.565,
  0.578, 0.589, 0.598, 0.606, 0.614). At high volume fractions there
  is a plateau on intermediate time scales. The final relaxation is
  slowed down strongly with increasing volume fraction, indicative of
  glassy dynamics.}
\end{center}
\end{figure}

To demonstrate the slowing down of the positional degrees of freedom, we
consider the self part of the intermediate scattering function,
$\Fsqt = \langle \exp[i {\bf q} \cdot \Delta{\bf r}(t)] \rangle$,
where ${\bf q}$ is the wave vector, $\Delta{\bf r}(t)$ the displacement of an
ellipsoid after time $t$, and the angle brackets denote average over particles
and ensemble average. In isotropic systems, $F_{\rm s}$ is a function of the
absolute value $q$ only. Decay of $\Fsqt$
indicates that structural relaxation has occured on the length scale set by 
$q$. 
In \Fig{fs} we show $\Fsqt$ for all simulations. The wave
length $q_{\rm max}$ was chosen close to the first maximum of the static
structure factor (prolate: $6.85/w$, oblate: $7.85/w$), 
i.e.~it corresponds to the neighbor spacing. One can clearly
see the development of a plateau with increasing volume fraction. This
means that there are two distinct stages of relaxation, and the latter slows
down dramatically upon increase of volume fraction (note
the logarithmic time scale). This
two-step decay is a typical phenomenon in glass formers
\cite{goetzeLiqFreezGlass:1991}. It is
interpreted in terms of particles being trapped in cages formed 
by their nearest neighbors. The initial decay corresponds to motion
within the traps, and the final decay to escape. At high volume
fractions, exceedingly cooperative rearrangements are required for
escape, making such events rare. Indicators of caging have not been
seen in hard ellipsoids before (unless the moment of inertia was strongly
increased \cite{Theenhaus.Allen.Letz.Latz.Schilling:2002}).


\begin{figure}
\begin{center}
\includegraphics[ 
  width=\figwidth
  ]{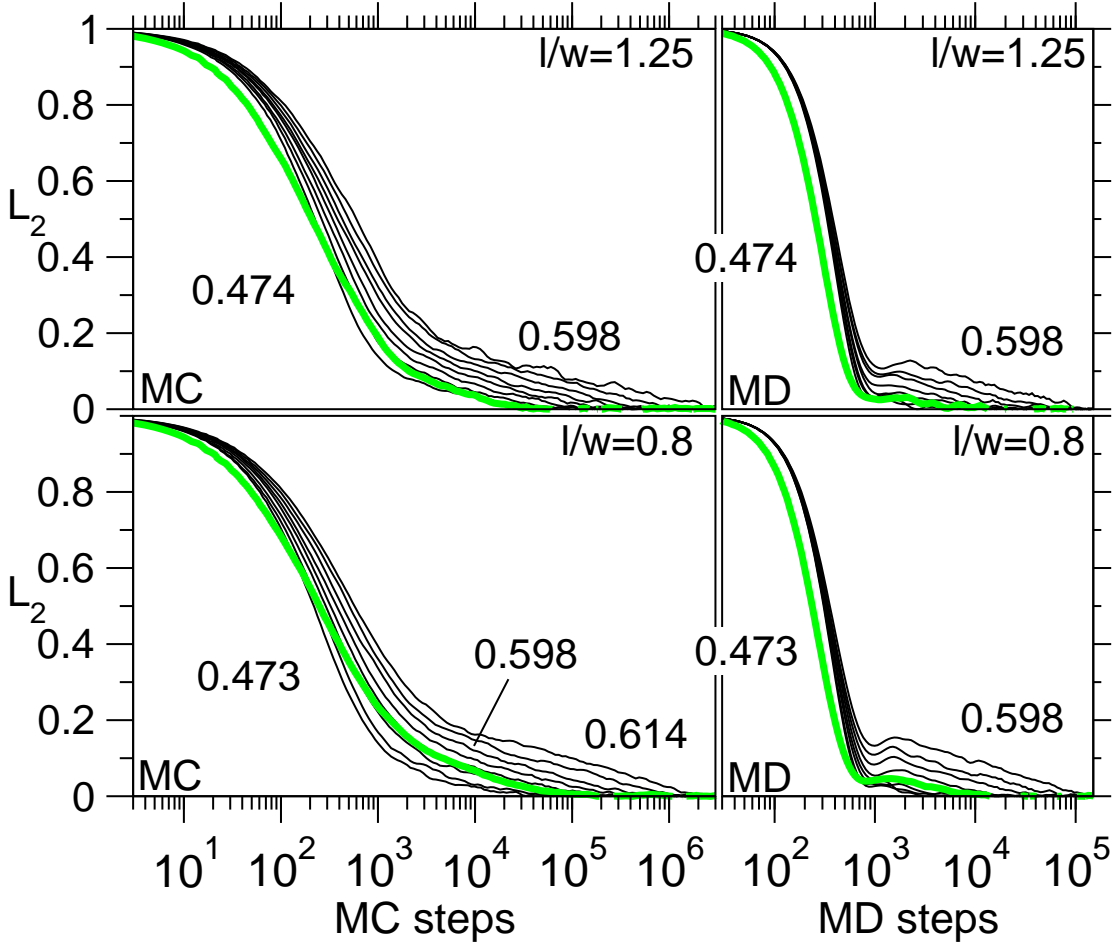}
\caption{\label{L2} (Color online) Second-order orientational
  correlators at several
  volume fractions $\phi$ as in \Fig{fs}. Again a plateau develops with
  increasing $\phi$. Hence, 
  orientational degrees of freedom are coupled to the positional ones. 
  Also shown are the third-order correlators at the highest density 
  (bold green lines). They do not slow down, indicating that
  flipping modes are not affected.} 
\end{center}
\end{figure}

Dynamics in the orientational degrees of freedom is observed in terms of the
second-order orientational correlator
$L_2(t) = (1/2) \langle 3 \cos^2\theta(t)-1 \rangle$, where $\theta(t)$ is the
angle between the orientation at time $t$ and the original orientation of an
ellipsoid. Decay of this function indicates that relaxation of orientational
degrees of freedom has taken place. Since $L_2(t)$ is an even function in
$\cos\theta(t)$, the head-to-tail symmetry of the particles is
taken into account. \Figlong{L2} shows the orientational correlation 
functions. As in the intermediate scattering functions, plateaus
develop at high volume fractions. Evidently, the
cages hinder rotations of the ellipsoids. As a
consequence, orientational and positional degrees of freedom are coupled. This
is in contrast with the decoupling found at lower volume fractions
\cite{demichele.schillingr.sciortino:2007}.

The shape of both positional and orientational correlators differs between MC
and MD on short 
time scales, reflecting the individual microscopic dynamics. On
intermediate and long time scales, they do not differ
significantly. Furthermore, when the correlators of the highest 
few volume fractions
are rescaled by their decay time, their long time parts fall onto a master
curve. These properties confirm predictions of MCT
\cite{Goetze.Sjoegren:1992}.

Unlike $L_2(t)$, the third-order orientational correlation function
$L_3(t) = (1/2) \langle 5 \cos^3\theta(t)- 3 \cos\theta(t) \rangle$
does not show plateaus (bold green lines in \Fig{L2}). Hence, while the 
overall reorientation slows down, flipping is
barely hindered. This is in accord with the MMCT prediction of Letz et
al.~\cite{letz.schilling.latz:2000} and has also been found for the case of
diatomic Lennard-Jones dumbbells \cite{Kammerer.Kob.Schilling:1997}, and
symmetric Lennard-Jones dumbbells
\cite{Chong.Moreno.Sciortino.Kob:PRL2005}. We note in passing that
crystallization, if it occurs, releases the orientational degrees of
freedom: The orientational correlators accelerate by three orders of magnitude
and no longer have a plateau.

\begin{figure}
\begin{center}
\includegraphics[ 
  width=\figwidth
  ]{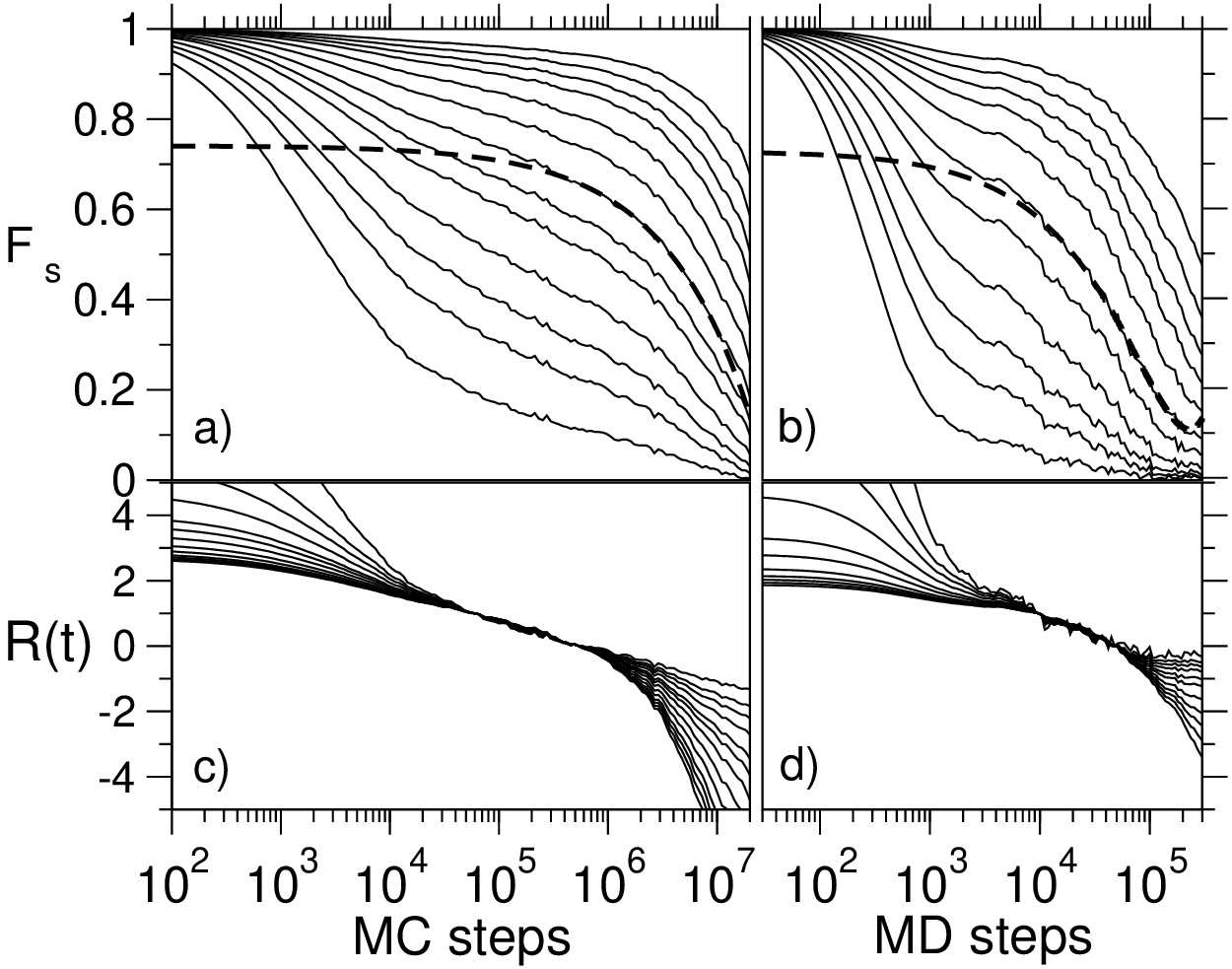}
\caption{\label{Fs_manyq} Self-intermediate scattering
  functions $\Fsqt$ of a) the oblate system (MC data) and b) the
  prolate system (MD data) at the 
  highest volume fraction, for the range $2.8 < qw < 20$. 
  The dashed lines show fits to the von-Schweidler law
  $\Fsqt = f_{\rm q} - h_{\rm q}^{(1)} t^b + h_{\rm q}^{(2)} t^{2b}$ with
  $b=0.65$. c) and d) show the same data transformed to
  $R(t) = [\Fsqt - F_{\rm s}(q,t_1)]/[F_{\rm s}(q,t_2) - F_{\rm
    s}(q,t_1)]$. The collapse of the functions onto master curves
  demonstrates the factorization property.
  }
\end{center}
\end{figure}

Next, we show that the slowing down in our systems is strong enough to
test MCT 
asymptotic laws. To this end, we first return to the intermediate scattering
functions and focus on their 
$q$-dependence at high volume fractions. \Figlong{Fs_manyq} presents these
functions for a) the oblate system (MC) and b) the prolate system (MD; in each
case the other method shows similar results), with wave
lengths in the range $2.8 < qw < 20$. We test two MCT predictions
for the vicinity of the glass transition \cite{goetzeLiqFreezGlass:1991}. 
Firstly, for the late stages of the
plateau and the early stages of the final decay, these functions should be
well-fitted by the von-Schweidler law (incl.~the
second-order correction), $\Fsqt = f_{\rm q} - h_{\rm q}^{(1)} t^b + h_{\rm
  q}^{(2)} t^{2b}$, where $f_{\rm q}$ is the plateau height, $h_{\rm q}^{(i)}$
are amplitudes, and $b$ is a system-universal exponent (also independent of
the microscopic dynamics). Agreement is excellent, as shown for two
examples in \Fig{Fs_manyq} (dashed lines). MC and MD data of both
systems are consistent
with $b=0.65 \pm 0.2$. Secondly, where the $\Fsqt$ are near their
plateaus, they should obey $\Fsqt = f_{\rm q} + h_{\rm q} G(t)$, where 
$h_{\rm q}$ is an amplitude, and $G(t)$ is
a system-universal function. This relation entails the ``factorization
property'', i.e.~that $\Fsqt$ can be factorized into a $q$-dependent and a
$t$-dependent part. To test this property for our systems, we transform $\Fsqt$
to $R(t) = [\Fsqt - F_{\rm s}(q,t_1)]/[F_{\rm s}(q,t_2) - F_{\rm s}(q,t_1)]$,
as done in \cite{gleim-epjb-13-83-2000}, where $t_1$ and $t_2$ are times in
the regime where 
the property holds. Since $R(t)$ is not a function of $q$, all
correlators should fall onto a single master curve. Moreover, the
curves should remain
ordered, i.e.~a curve which is above another on the left-hand side remains
above the other on the right-hand side. Panels c) and d) of
\Fig{Fs_manyq} demonstrate the validity of the factorization property,
and indeed they remain ordered.


\begin{figure}
\begin{center}
\includegraphics[ 
  width=\figwidth
  ]{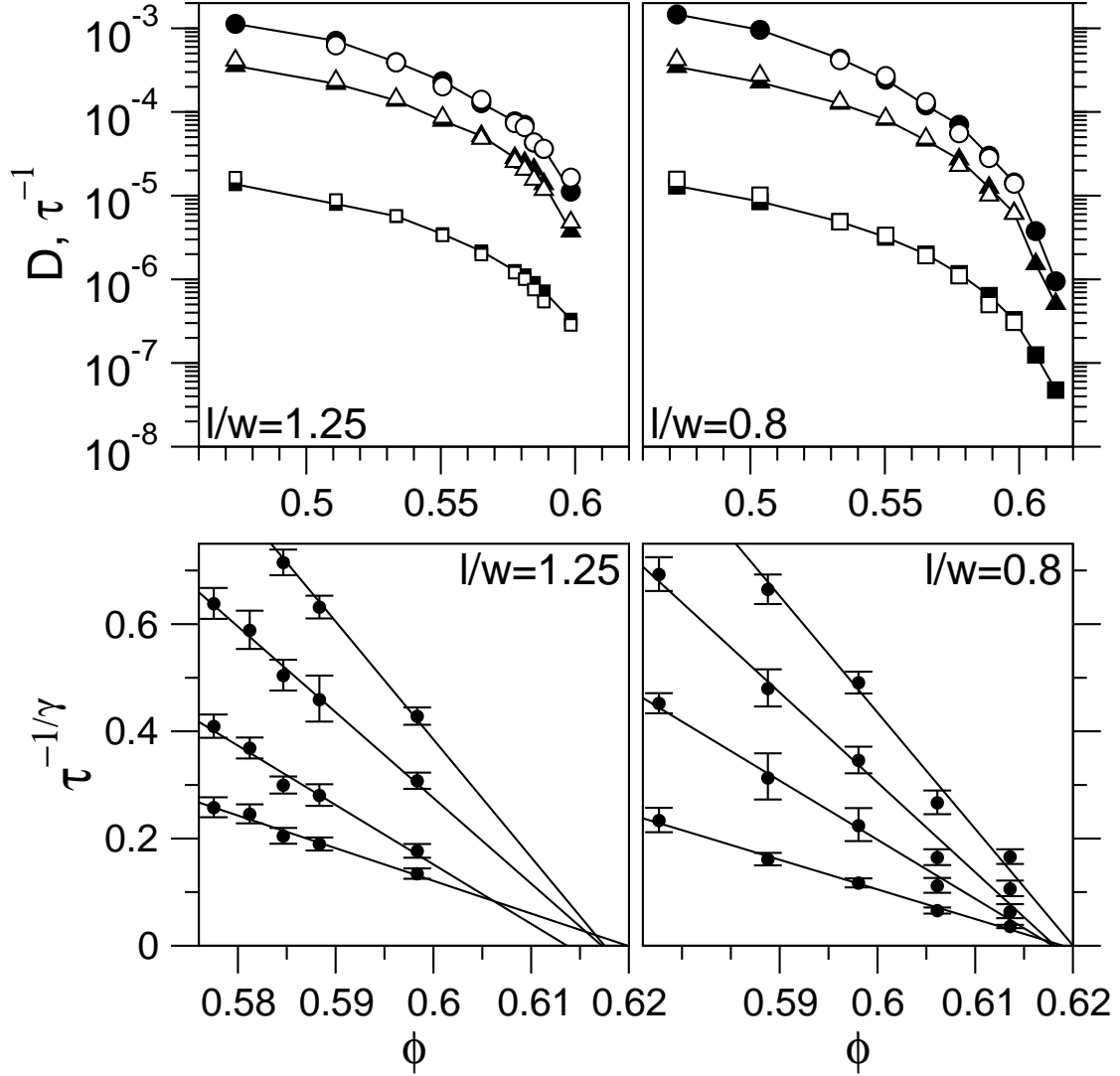}
\caption{\label{relax_all} Upper panels: Inverse relaxation times
  $\tau^{-1}$ obtained 
  from $F_{\rm s}(q_{\rm max},t)$ (triangles) and from $L_2 (t)$ (circles),
  and diffusion constants $D$ (squares), as a function of volume fraction
  $\phi$. MC data
  (filled symbols, lines to guide the eye, and rescaled to match MD time scale)
  and MD data (open symbols) show 
  excellent agreement. Lower panels (MC data only): $\tau^{-1/\gamma}$
  multiplied by arbitrary factors for clarity, from
  $\Fsqt$ (several values of $q$) and $L_2(t)$, demonstrating the validity of
  the MCT scaling law $\tau^{-1} \propto (\phi_c - \phi)^{\gamma}$. 
  $\gamma = 2.3$. The
  straight-line fits indicate a glass transition volume fraction of
  $\phi_c = 0.615 \pm 0.005$ (prolate system) and
  $\phi_c = 0.618 \pm 0.005$ (oblate system). MD data agree.
}
\end{center}
\end{figure}

We finally demonstrate the applicability of an MCT scaling law for the
slowing down, which even allows us to extract the MCT glass transition volume
fraction.
\Fig{relax_all} shows relaxation times and diffusion constants as a
function of volume fraction. The relaxation times are defined as the times
at which the positional correlators have reached the value 0.1, and the
orientational correlators have reached the value 0.02 (since the plateau is
quite low). The diffusion constants were determined from the mean squared
displacements via the Einstein relation
$D=\lim_{t\to\infty} \frac{d}{dt} \langle (\Delta{\bf r}(t))^2 \rangle/6$.
The upper panels of \Fig{relax_all} show inverse relaxation times $\tau^{-1}$ 
obtained from $F_{\rm s}(q_{\rm max},t)$ (triangles) 
and from $L_2 (t)$ (circles),
and diffusion constants $D$ (squares). 
The MC values (filled symbols) have been multiplied by a factor of 24
(prolate) and 18 (oblate)
to match the MD time scale ($L_2$: 19 and 16, respectively).
The errors are of symbol size or
less. Note that in each system, one common factor for the positional
variables yields 
excellent agreement between MC and MD (open symbols), in agreement with
MCT. The factor for the orientational relaxation times need not be the same
since it depends on the choice
of the orientational MC move size. The slowing down of all variables is
super-Arrhenius. According to MCT, it should approach a power law of the form
$D \propto \tau^{-1} \propto (\phi_c - \phi)^{\gamma}$, where $\phi_c$ is the
MCT glass-transition volume fraction, and $\gamma$ is related to the
von-Schweidler exponent $b$. Both $\phi_c$ and $\gamma$ should be
system-universal. In the lower panels of \Fig{relax_all} we
demonstrate the validity of this 
prediction for the MC results of $\Fsqt$ (from top: prolate: $qw$ = 6.85, 11, 16;
oblate: $qw$ = 7.85, 16, 20) and $L_2$
(bottom). The exponent $\gamma=2.3$ was chosen in agreement with 
$b=0.65$. The straight-line fits comply with a
common $\phi_c = 0.615 \pm 0.005$ for the prolate system and with
$\phi_c = 0.618 \pm 0.005$ for the oblate system.
The fact that there is a common value 
for positional and orientational relaxation times further demonstrates 
the strong coupling of these
degrees of freedom. We found agreement with the analogous
analysis of the MD data. The values we found differ from the MMCT
predictions of Letz et al.~\cite{letz.schilling.latz:2000}, 
viz.~$\phi_c = 0.540$ and $0.536$ ($l/w = 1.25$ and $0.80$,
respectively). The mismatch between the numerical MCT calculations based on
static structure, and scaling law fits based on simulated dynamics is, however,
not unusual \cite{kob-pre-52-4134-1995,nauroth-pre-55-657-1997} and has been
attributed to activated (``hopping'') processes for
which MCT does not account. A similar mismatch is found in the hard-sphere
system \cite{Voigtmann.Puertas.Fuchs:2004}. We note that the present
study also displays the prolate-oblate symmetry seen in previous work
on the equilibrium properties and dynamics of ellipsoids
\cite{frenkel.mulder:1985,letz.schilling.latz:2000,demichele.schillingr.sciortino:2007}. However,
we observed that crystallization does not have this symmetry: the
prolate system crystallizes more readily.

In summary, we have performed molecular dynamics and Monte Carlo
simulations of the hard-ellipsoid fluid. In this very simple anisometric model
we observe glassy dynamics sufficiently strong that MCT asymptotic scaling 
laws can be tested and are found to apply. 
We find strong coupling of positional and orientational
degrees of freedom, leading to a common value for the glass-transition volume
fraction $\phi_c$ for positional and
orientational relaxation times ($l/w = 1.25$: $\phi_c = 0.615 \pm 0.005$,
$l/w = 0.80$: $\phi_c = 0.618 \pm 0.005$).
The presence of glassy dynamics has been predicted by MMCT. 
However, as MMCT cannot make a statement about crystallization, a test 
by simulation was required. We argue that
particle anisometry acts as a sufficient source of disorder to prevent
crystallization. This sheds new
light on the question of which ingredients 
are required for glass formation. 
Experimental studies of glassy dynamics in
the isotropic phase of 
liquid crystals have been conducted \cite{cang-jcp-118-9303-2003}, but 
not in ellipsoids. It is possible to synthesize ellipsoids of colloidal size
\cite{keville.etal:1991,ho.etal:1993} and to study their dynamics with
confocal microscopy \cite{mukhija-jcis-314-98-2007}. In the light of the 
above, an experimental study of glassy dynamics in colloidal hard ellipsoids 
seems very promising.

This work was supported by the Emmy Noether Program and SFB TR6 of the
Deutsche Forschungsgemeinschaft (DFG), and the European Network of Excellence 
SoftComp. We are grateful to M.~P.~Allen for sharing his molecular
dynamics code, to the NIC J\"ulich for computing time, and to K.~Binder,
R.~Schilling, M.~P.~Allen, J.~Horbach, J.~Baschnagel, W.~Kob, M.~Letz and
W.~A.~Siebel for helpful suggestions. 

\bibstyle{revtex}

\end{document}